\documentclass{article}

\usepackage{arxiv}

\usepackage[utf8]{inputenc} 
\usepackage[T1]{fontenc}    
\usepackage{hyperref}       
\usepackage{url}            
\usepackage{booktabs}       
\usepackage{amsfonts}       
\usepackage{nicefrac}       
\usepackage{microtype}      
\usepackage{lipsum}
\usepackage{graphicx}
\graphicspath{ {./images/} }
\usepackage{tfrupee}

\title{Unlocking AI’s Potential in Agriculture: The Critical Role of Data}

\author{
    K. B. Vedamurthy \\
    Dept. of Dairy Business Management, Dairy Science College, KVAFSU, Bengaluru, India \\
    \texttt{vedandri@gmail.com} \\
    \And
    Manojkumar Patil \\
    Dept. of Computer Science and Automation, IISc, Bengaluru, India \\
    \And
    Vaishnavi \\
    Dept. of Dairy Business Management, Dairy Science College, KVAFSU, Bengaluru, India \\
    \And
    Priyanka V \\
    Dept. of Computer Science and Automation, IISc, Bengaluru, India\\
    \And
    Suman L \\
    Dept. of Agricultural Economics, UAS, Bengaluru, India\\
    \And
    Ajayakumar \\
    Dept. of Agricultural Economics, UAS, Bengaluru, India \\
    \And
    Sagar \\
    Dept. of Dairy Business Management, Dairy Science College, KVAFSU, Bengaluru, India \\
}

\begin{document}
\maketitle
\begin{abstract}
India generates substantial volumes of public agricultural data, yet artificial intelligence (AI) adoption in farming remains limited and largely confined to pilot initiatives. This paper examines this gap by assessing India’s agricultural data infrastructure against the requirements of AI systems deployed at scale. Drawing on a systematic review of major national datasets and digital initiatives—including Soil Health Cards, crop insurance, AgriStack, and selected state platforms—we identify persistent structural constraints, including temporal misalignment between data collection and agricultural decision cycles, spatial fragmentation arising from the absence of common geocodes linking soil, weather, and yield information, limited machine readability due to reliance on static data formats, and unclear governance frameworks that restrict data access and reuse. These deficiencies impede cross-dataset integration and automated decision support, with disproportionate consequences for smallholders, who constitute 86~\% of India’s farmers and lack the capacity to compensate for weak data infrastructure. Drawing on implementation evidence from India and comparative international experiences, the paper identifies recurring features associated with scalable digital agriculture systems, including incentives linked to data provision, service bundling through local institutions, and sensor-enabled risk management. Building on these insights, we define the core attributes of AI-ready agricultural data—persistent identifiers, machine-accessible formats, temporal alignment, semantic interoperability, and transparent governance—and show that even India’s most advanced initiatives remain only partially aligned with these requirements. The paper’s contribution is primarily diagnostic: it demonstrates that data infrastructure limitations, rather than algorithmic capability, constitute the principal bottleneck to scalable AI adoption in Indian agriculture and proposes an operational for assessing agricultural data readiness.
\end{abstract}


\section{Introduction}\label{sec1}

India has built some of the world’s most ambitious digital public infrastructures, yet the diffusion of Artificial Intelligence (AI) in agriculture remains strikingly limited. While AI adoption in sectors such as healthcare, finance, and manufacturing in India exceeds 25--40~\%, agricultural applications remain fragmented and largely pilot-driven, despite the sector engaging nearly 45.5~\% of the national workforce \cite{cfo_et_ai_healthcare_2025}. This contrast presents a central puzzle: why has AI scaled rapidly in other data-intensive sectors but not in agriculture?

Evidence from sectors where AI adoption has accelerated most rapidly suggests a common enabling condition: data standardization, enforced either through regulation or through dominant digital platforms. In agriculture, neither mechanism operates effectively. Agricultural data remain dispersed across institutions, collected at incompatible spatial and temporal scales, and governed by fragmented mandates. Access to data and systematic integration across sources are largely absent, while persistent concerns regarding data quality further undermine analytical reliability.

These limitations are particularly consequential for data-driven policymaking, planning, and monitoring, where inconsistent or non-digitized data constrain evidence-based decision-making. Although data generation occurs at multiple administrative and operational levels, it is rarely digitized in a standardized manner or integrated across systems. As a result, constraints in data availability, quality, and interoperability—rather than algorithmic capability—have emerged as the primary bottlenecks limiting the scalable deployment of AI in Indian agriculture.

Although recent discourse on digital agriculture emphasizes AI-enabled applications such as yield forecasting, pest detection, and market intelligence, \textbf{data systems} are often treated as a secondary concern. This review argues that agricultural data infrastructure, encompassing data generation, integration, governance and interoperability, constitutes the primary bottleneck to scaling AI in agriculture. In smallholder-dominated systems such as India’s, where 89.4~\% of agricultural households operate less than two hectares of land \cite{pib2025a}, these constraints have direct implications for equity, as failure to scale data-driven tools risks excluding the very farmers whom food security and rural development policies seek to support.

The paper advances a data-first perspective that conceptualizes agricultural data systems not as passive inputs, but as active infrastructure shaping the feasibility, scalability and inclusiveness of AI deployment. It provides a systematic assessment of India’s emerging public agricultural data initiatives, including Agri Stack, UPAg and related platforms, against the technical and institutional prerequisites of production-grade AI, while explicitly acknowledging their developmental constraints and adoption timelines \cite{nceg_27th_compendium}. Drawing on international experiences, the review identifies persistent structural misalignments and proposes implementable strategies to strengthen agricultural data ecosystems \cite{weforum_beyond_data_2025}. By foregrounding data infrastructure as a binding constraint, the paper contributes to policy and academic debates on how digital agriculture can realistically support productivity growth, climate resilience, and inclusive development in India. This paper concentrates on the data-infrastructure layer of digital agriculture; AI applications are discussed only to the extent that they reveal data gaps, governance failures, or scalability constraints.

\subsection{Background}

The Green Revolution marked a major structural transformation of Indian agriculture, enabling food self-sufficiency through state-led interventions that promoted high-yielding varieties, chemical fertilizers, irrigation expansion and institutional support systems \cite{hans2024revisiting}. During this period, cereal production nearly tripled, even though cultivated land expanded by only about 30~\% \cite{pingali}. This input-intensive model fundamentally reshaped India’s agrarian economy and ensured national food security.

In the decades following the Green Revolution, however, Indian agriculture has not undergone a transformation of comparable scale. Despite contributing approximately 18.20~\% to national GDP and supporting 42.30~\% of the population, the sector continues to face persistent challenges, including low productivity growth, increasing climate vulnerability, fragmented landholdings, and declining factor efficiency \cite{economic_survey_2025}. These structural constraints have intensified the search for new pathways to agricultural modernization.

Recent policy initiatives and technological advances have increasingly emphasized data-driven and digitally enabled agriculture as a potential source of renewed productivity growth. In India, the economic potential of data-enabled agriculture is estimated at nearly \$65 billion, yet the share of this value accruing to small and marginal farmers remains limited \cite{wef_future_farming_2025}. Advances in sensing technologies, remote sensing, mobile platforms, and cloud computing have expanded the scale and granularity of agricultural data generation. Nevertheless, the realization of AI-driven agriculture remains limited, as most algorithms depend on large volumes of accurate, georeferenced, and temporally consistent data that are not yet systematically available.

Parallel advances in artificial intelligence have delivered substantial economic impacts in non-agricultural sectors, prompting interest among policymakers, agri-tech firms and extension systems in applying AI to agriculture. Yet, the translation of these technologies into broad-based agricultural impact remains uneven. This review argues that these limitations arise less from deficiencies in algorithms and more from weaknesses in the underlying data systems that support AI deployment.

While data is widely acknowledged as essential for AI, this paper advances the literature by conceptualizing agricultural data systems as active infrastructure that shapes the feasibility, scalability, and equity of digital agriculture. By examining India’s public agricultural data initiatives alongside international experiences, the paper identifies structural misalignment between existing data architectures and the requirements of AI-driven applications. It shows that weaknesses in data standardization, interoperability and governance directly affect model reliability, spatial transferability and inclusiveness, making data system reform a necessary condition for aligning digital agriculture with India’s broader growth objectives \cite{coingeek_ai_growth_niti_2025}.

\subsection{Outline of the Paper}

This paper provides a systematic analysis of agricultural data infrastructure challenges and opportunities for AI deployment in Indian agriculture, progressing from existing conditions to actionable pathways for reform.

Section~\ref{sec2} establishes the foundation by examining India’s agricultural data landscape, cataloguing major data sources across government ministries and agencies and identifying key governance challenges, including fragmentation, inconsistent identifiers and state-level policy variation that impede seamless integration.

Section~\ref{sec3} grounds the analysis in implementation realities by examining recent digital agriculture initiatives and field deployments, documenting technical and operational challenges related to data consistency, validation mechanisms, update lags and user adoption.

Section~\ref{sec4} systematically analyzes barriers to effective data utilization across three dimensions: data fragmentation and integration failures, quality and accessibility constraints and institutional limitations that restrict interoperability and real-time analytics.

Section~\ref{sec5}\label{sec5}, \textit{What Makes Agricultural Data AI-Ready?}, moves beyond diagnosis to definition. It addresses a foundational question for digital agriculture infrastructure: what data characteristics are required for scalable AI deployment? The section defines AI-ready agricultural data in terms of standardization, temporal relevance, spatial alignment, machine accessibility and governance conditions that enable direct integration into AI workflows without extensive pre-processing.

Section~\ref{sec6} evaluates ongoing public digital infrastructure reforms, assessing their development status, pilot-stage limitations, API readiness and projected deployment timelines, with particular attention to technical specifications and scalability constraints.

Section~\ref{sec7} examines private sector engagement, detailing data access constraints faced by technology firms, adaptive strategies used to operate within these limitations and the influence of state-level protocols on public–-private collaboration.

Section~\ref{sec8} explores AI application domains that become feasible under robust data systems, including precision agriculture, predictive analytics, resource optimization and market intelligence, while emphasizing that application performance is conditional on upstream data quality and governance.

Section~\ref{sec9} draws lessons from international experiences across developed and developing contexts, analyzing both successful implementations and instructive failures. The comparative analysis extracts policy and technical insights relevant to India’s smallholder-dominated agriculture and diverse agro-climatic conditions.

Section~\ref{sec10} synthesizes the review’s findings into actionable implementation strategies, outlining phased interventions spanning immediate technical fixes, medium-term institutional reforms, and long-term ecosystem development. Emphasis is placed on interoperable public APIs, federated learning architectures, and sustainable public–private partnership models.

Section~\ref{sec11} concludes by distilling key insights and identifying priorities for future research and policy action, reaffirming the central role of agricultural data infrastructure in enabling scalable, inclusive AI deployment.

\section{India's Agricultural Data}\label{sec2}

\subsection{Data Landscape}\label{subsec:landscape}
India's agricultural data ecosystem comprises extensive but fragmented datasets spanning crop statistics, meteorology, market dynamics, and land records. While the volume of generated data is substantial, critical gaps in temporal cadence, spatial resolution, and public accessibility currently impede the deployment of production-grade AI.

\begin{table}[h!]
\caption{Major Public Agricultural Data Sources in India}
\label{tab:agri_data_sources}
\small
\centering 
\begin{tabular}{p{3cm} p{2.5cm} p{2.5cm} p{3.6cm}} 

\toprule
\textbf{Data Type} &
\textbf{Agency} &
\textbf{Portal} &
\textbf{Update \& Access} \\
\midrule
Crop area, production, yield &
MoA\&FW &
\url{www.desagri.gov.in} &
Annual / seasonal; Public \\

Market prices, arrivals (Agmarknet) &
DMI (MoA\&FW) &
\url{www.agmarknet.gov.in} &
Daily; Public \\

Weather observations &
IMD &
\url{www.mausam.imd.gov.in} &
Hourly / daily; Public (aggregated) \\

Fertilizer consumption and subsidy &
DoF &
\url{www.fert.gov.in} &
Monthly / quarterly; Limited \\

Minimum Support Prices (MSP) &
CACP &
\url{www.cacp.da.gov.in} &
Annual (pre-season); Public \\

Agricultural census &
DA\&FW &
\url{www.agcensus.gov.in} &
5-yearly; Public \\

Water resources and irrigation &
Jal Shakti/CWC &
\url{www.indiawris.gov.in} &
Daily / weekly; Public \\

Digital land records &
MoRD / States &
\url{www.bhunaksha.nic.in} &
Irregular; Limited \\

Satellite-based crop monitoring &
ISRO / MNCFC &
\url{www.ncfc.gov.in} &
Seasonal; Restricted \\

Agri export--import statistics &
APEDA &
\url{www.apeda.gov.in} &
Monthly / annual; Public \\

Merchandise trade statistics &
DGCI\&S &
\url{www.dgciskol.gov.in} &
Monthly; Public \\

District level Crops, livestock, infrastructure, census, GDP, environment and prices & ICRISAT & \url{http://data.icrisat.org/dld/} & Annual \\

International trade flows &
UN Comtrade/ITC Trademap &
\url{www.comtrade.un.org}/\url{trademap.org} &
Annual / quarterly; Public \\

Trade regulation and Non Tariff Measures (NTMs) & TRAINS online (UNCTAD) & \url{https://trainsonline.unctad.org/home} &  Monthly updates \\

Food and agriculture statistics (FAOSTAT) &
Food and Agriculture Organization &
\url{www.fao.org/faostat} &
Annual (some indicators updated continuously); Public \\

\bottomrule
\end{tabular}
\vspace{2mm}

\footnotesize{Source: Author’s compilation from official national and international data portals.}
\end{table}

Table~\ref{tab:agri_data_sources} catalogs the primary public agricultural datasets. While weather data provides high-frequency updates, it lacks the farm-level granularity required for precision interventions. Conversely, market price data—--though updated daily---is hampered by inconsistent commodity classifications across state jurisdictions. Crop statistics and census data, while methodologically rigorous, operate on annual cycles-a frequency insufficient for real-time predictive modeling. Furthermore, high-resolution satellite monitoring remains largely restricted to government agencies, creating a `data silos' effect that stifles private-sector AI innovation.

\subsubsection*{AI-Ready Assessment Framework} 
To support scalable AI, agricultural datasets must satisfy five core technical benchmarks derived from current digital public infrastructure standards \cite{Fu2025AIReady,GoM2025MahaAgriAI}:
\begin{enumerate}
    \item \textbf{Low Temporal Latency:} Data delivery cadence must align with agricultural decision cycles (e.g., real-time weather or weekly pest alerts) to ensure model relevance \cite{Fu2025AIReady}.
    \item \textbf{Fine Spatial Granularity:} High-resolution datasets enabling plot-level analysis are essential for precision-scale interventions.
    \item \textbf{Semantic Interoperability:} Adherence to standardized taxonomies and machine-actionable metadata is required for automated cross-domain data discovery \cite{Fu2025AIReady}.
    \item \textbf{Predictable Governance:} Transparent protocols for data access, privacy compliance (e.g., DPDP Act), and licensing are critical for ethical AI scaling \cite{GoM2025MahaAgriAI}.
    \item \textbf{Ground-Truth Validation:} Datasets must include robust field-level verification mechanisms to minimize algorithmic bias and refine model accuracy \cite{Fu2025AIReady}.
\end{enumerate}

At present, no Indian agricultural dataset satisfies all five criteria. While the Digital Crop Survey (DCS) and AgriStack represent significant progress—mapping 253 million plots and issuing 60 million Farmer IDs as of 2024-25—these initiatives remain in pilot stages \cite{ICRISAT_AgriStack_2025}. Limited API access for external developers persists as a primary bottleneck, as researchers face opaque approval processes and non-standardized data-sharing agreements that discourage the transition from academic pilots to commercial deployments \cite{wef_future_farming_2025}.

\subsection{Governance Constraints on AI Integration}\label{subsec:governance}

Institutional fragmentation constitutes a significant barrier to the multi-source data integration required for complex AI applications \cite{coble2018big}. Agricultural datasets are maintained by multiple ministries operating under distinct mandates and technical standards, limiting interoperability. In particular, the absence of consistent national identifiers for farms and crops constrains the automated linkage of soil, weather, and yield data necessary for integrated agronomic analysis.

Governance heterogeneity across states further complicates nationwide deployment. While some states have introduced more standardized digital procedures, others continue to rely on discretionary approval processes. As a result, AI developers often face the need to negotiate data access separately across jurisdictions, reducing scalability and increasing transaction costs.

Farmer data ownership and consent frameworks also remain underdeveloped in operational terms. Although recent data protection legislation has strengthened legal safeguards, practical guidance on consent architectures and implementation remains limited. In the absence of standardized, machine-readable consent mechanisms, granular farmer-level data is frequently underutilized, constraining the continuous ground-truth validation required for reliable AI performance \cite{wef_future_farming_2025}.

The Digital Agriculture Mission (allocated \rupee 2,817 crore) seeks to bridge these gaps via AgriStack’s consent-driven architecture \cite{PIB_AgriStack_2024}. However, with nationwide deployment scheduled through 2026-27 and only 19 states currently onboarded via MoUs, the infrastructure remains in a transitional state. Until these governance frameworks reach maturity, India’s data ecosystem will remain only partially aligned with the requirements of production-grade AI.

\section{Implementation Experience: Lessons from National Agricultural Programs}\label{sec3}

India’s agricultural digitization initiatives reflect strong policy intent and sustained public investment. Field-level implementation experience, however, indicates that outcomes often diverge from design expectations due to limitations in underlying data systems. This section examines two nationally implemented programs---the Soil Health Card (SHC) scheme and the Pradhan Mantri Fasal Bima Yojana (PMFBY)-to illustrate how gaps in identifiers, metadata, validation and data access constrain the effectiveness and scalability of data-driven interventions.

\subsection{Soil Health Card Scheme (SHC)}

The SHC scheme, launched in 2015, aimed to deliver plot-specific soil diagnostics and fertilizer recommendations. While farmer awareness of the program is high, adoption of recommended fertilizer practices remains uneven. Based on a national sample of 3,600 farmers, only 48~\% reported following the printed fertilizer rates, despite 82~\% awareness of the scheme \cite{reddy}.

A key constraint lies in data design. Soil samples are aggregated over spatial grids covering multiple farmers, resulting in the loss of persistent plot-level identifiers and weakening parcel-level traceability. Although GPS coordinates are collected during sampling, outputs are typically delivered as static PDF cards. The SHC portal exposes no machine-readable JSON, CSV or API endpoints, requiring downstream advisory systems to manually extract or re-enter values, thereby limiting integration with AI-based nutrient optimization models.

Operational delays further reduce relevance. Laboratory capacity constraints and manual data entry practices extend turnaround times beyond critical crop decision windows. Adoption improves significantly when extension workers explain recommendations highlighting that interpretability and institutional mediation are essential complements to technical accuracy. The SHC experience demonstrates that data-driven advisories require persistent identifiers, standardized metadata, timely delivery, and machine-accessible formats to support sustained use.

\subsection{Pradhan Mantri Fasal Bima Yojana (PMFBY)}

PMFBY (2016) was designed as a technology-enabled crop insurance program incorporating remote sensing, mobile applications, and digital yield reporting. In practice, implementation continues to rely heavily on conventional Crop Cutting Experiments (CCEs), with digital capture and technology-based yield estimation systems, including YES-TECH, only partially implemented \cite{Ghosh,niti2016agri,LS_PMFBY_2036_2025}.

As a result, granular, georeferenced, and machine-actionable yield data at the village or insurance-unit level remain limited. Continued dependence on manual CCEs and incomplete spatial linkage between yield observations and cadastral units constrain reconciliation with satellite-derived yield rasters, weakening validation and actuarial accuracy.

Institutional coordination further conditions outcomes. Claim settlement timelines depend on synchronized data flows among state governments, insurers, and banks, and delays arise when required datasets cannot be generated, validated, or exchanged within prescribed time frames \cite{Ghosh}. Taken together, these experiences illustrate that advances in sensing and analytics cannot, by themselves, compensate for persistent gaps in identifiers, validation pipelines, and interoperable data architecture.

\subsection{Synthesis: Pathways Toward Data-Centric Maturity}

The implementation experiences of SHC and PMFBY indicate that gaps in outcomes are less a consequence of operational execution than of underlying data architecture. Rather than isolated programmatic shortcomings, these experiences reveal structural design choices that shape the scalability and reliability of AI-enabled interventions. Three priority areas for refinement emerge:

\begin{itemize}
    \item \textbf{Identity Resolution:} Moving from grid-based sampling to persistent, georeferenced plot identifiers (such as proposed AgriStack IDs) is critical for parcel-level traceability and longitudinal analysis \cite{GoM2025MahaAgriAI}.
    \item \textbf{Automated Validation:} Transitioning from manual crop-cutting experiments to AI-enabled remote sensing and data fusion pipelines can standardize yield validation and reduce actuarial delays \cite{GoM2025MahaAgriAI}.
    \item \textbf{Machine-Actionable Access:} Shifting from static reporting formats (e.g., PDFs) to APIs and machine-readable standards (JSON/CSV) enables the integration of public datasets into real-time AI advisory and risk-assessment workflows \cite{Fu2025AIReady}.
\end{itemize}

Together, these pathways frame data-centric maturity not as an expansion of data openness, but as a transition toward interoperable, verifiable, and continuously usable digital public infrastructure.

\section{Principal Barriers to AI Implementation}\label{sec4}

\subsection{Fragmentation and Institutional Silos}

Digital agriculture initiatives in India are frequently characterized by fragmented efforts operating across institutional silos, resulting in limited coordination, duplication of data collection, and constrained reuse \cite{worldbank2021}. State-level platforms such as Telangana’s AdeX illustrate how datasets on soil health, crop conditions, and farm practices are often generated using heterogeneous collection methods and spatial references, preventing seamless integration across programs. These mismatches restrict real-time analytics and frequently necessitate manual aggregation, delaying decision-making for farmers and policymakers \cite{datatopolicy}. Differences in system design and data standards further limit interoperability and the ability to generate comparable insights across regions \cite{Hong,idh2024unlocking,oecd2022}.

\textbf{Case Illustration: The Geocoding Gap}

Fragmentation in India’s agricultural data infrastructure is observable at the spatial level: Soil Health Card (SHC) grid polygons and PMFBY insurance-unit boundaries do not share a common geocode. The absence of a shared spatial identifier prevents systematic linkage of soil fertility information with yield-loss estimation workflows, constraining integrated analysis across soil, weather, and insurance datasets.

\subsection{Data Quality and Accessibility Constraints}

Persistent challenges related to agricultural data quality arise from inconsistent collection practices, limited validation mechanisms, and uneven digitization, particularly at disaggregated levels such as villages and plots \cite{datatopolicy}. Significant portions of administrative and advisory information continue to be managed through paper-based or partially digitized systems, limiting reliability for analytical and predictive use. Data accessibility is further constrained by cost barriers and restrictive sharing arrangements, which limit reuse by researchers, startups, and local innovators \cite{datatopolicy,idh2024unlocking,nitiaayog2018ai}. In parallel, uneven access to digital tools among smallholders reduces the reach and effectiveness of data-driven advisory services \cite{Aranguri,fao2023}.

\textbf{Case Illustration: The Verification and Access Penalty}

Field-level observation indicates gaps between printed Soil Health Card recommendations and analytically verifiable values in some cases. In addition, the absence of public APIs for accessing advisory data forces technology providers to rely on OCR-based digitization of PDFs, imposing substantial fixed costs and creating barriers to entry for smaller firms.

\subsection{Institutional Gaps and the Stewardship Vacuum}

Institutional constraints further limit AI scalability, including the absence of widely adopted interoperability standards, metadata conventions, and application programming interfaces (APIs) that would enable routine data exchange across organizations \cite{ambalavanan2025,nitiaayog2018ai}. The literature highlights weak coordination mechanisms and unclear assignment of responsibility for data stewardship across agencies as persistent governance challenges \cite{alalawneh2020,gupta2022}. Without clearly defined custodianship and harmonized protocols, data systems remain difficult to align and sustain at scale.

\textbf{Case Illustration: Temporal Reporting Lags}

The absence of a unified, machine-readable identifier for crop-sown information contributes to delays in data consolidation. In the case of PMFBY, yield and enrollment data often reach central systems several weeks after harvest, limiting their usefulness for time-sensitive AI applications such as in-season risk assessment or pest and weather stress monitoring.

\subsection{Structural Limits}

Even where agricultural data are formally digitized, they are frequently not machine-consumable \cite{su142316131}. Foundational agronomic knowledge remains embedded in unstructured documents, creating a final structural barrier to AI deployment.

\textbf{Case Illustration: The Package of Practices}

Crop Package of Practices documents are typically released as static PDFs without standardized crop or location identifiers. As a result, agronomic logic cannot be directly parsed or queried by automated systems, rendering these documents unsuitable for plot-level AI models without extensive manual pre-processing.

\subsection{Semantic Ambiguity and the Linguistic Gap}

Even when agricultural data are digitized, they often lack a shared ontological framework. In the Indian context, the same biological entity—such as a crop, pest, or weed—may be referenced using different vernacular names across states or even neighboring districts\cite{drury2019semantic}. In the absence of a semantic mapping layer that links localized terminology to standardized scientific taxonomies, AI systems remain geographically bounded. This semantic fragmentation limits transfer learning, constraining the reuse of models trained in one agro-climatic zone when applied to another.

\textbf{Case Illustration: The Linguistic Gap}

Agricultural advisory and surveillance datasets frequently encode pest observations using localized names. For example, Fall Armyworm may be referred to by different vernacular terms across regions. Without a linked-data approach that maps these local labels to a common taxonomic identifier, AI models trained on one dataset may fail to interpret equivalent observations from another region, despite the underlying agronomic threat being identical.

\subsection{Data Sovereignty and the Trust Gap}

Beyond technical considerations, data sovereignty and ownership frameworks play a critical role in shaping AI readiness. In many smallholder-dominated systems, uncertainty regarding how plot-level data are stored, shared, and monetized reduces farmer willingness to participate in digital data ecosystems \cite{SULLIVAN2024100477,yeo}. When governance arrangements do not clearly articulate reciprocal benefits or usage rights, AI developers face constraints in accessing consistent, high-quality ground-truth data required for model calibration and validation.

\textbf{Case Illustration: The Trust Gap}

Several digital agriculture platforms function primarily as unidirectional data pipelines, collecting field-level information from farmers without providing transparent records of ownership, reuse, or downstream value generation. This asymmetry can reduce incentives for careful data entry and sustained participation, resulting in noisier inputs and limiting the reliability of AI models that depend on farmer-contributed data.

\section{Defining AI-Ready Agricultural Data}\label{sec5}

Earlier sections showed that fragmented identifiers, delayed updates, and non-machine-readable formats constrain AI deployment in Indian agriculture. This section therefore addresses a practical question: what constitutes \emph{AI-ready} agricultural data?
We define AI-ready data as structured, standardized, and machine-accessible datasets that can be ingested into AI pipelines with minimal manual pre-processing or ad hoc reconciliation \cite{JonkerIBM2025,johnk2021ready}.

We use crop Package of Practices (PoP) \cite{nfsmpop} documents as an illustrative case. PoPs directly inform advisory services and precision agriculture applications, yet currently exist almost exclusively as PDFs or unstructured web pages, limiting their suitability for automated reasoning, optimization, or model integration.

\subsection{From Documents to AI-Consumable Data}

PoP documents typically span 40-60 pages and combine narrative text, tables, and conditional advice. While suitable for human interpretation, this format introduces systematic barriers for AI use \cite{aiready}, including inconsistent terminology, mixed units, ambiguous timing, and the absence of programmatic access. Table~\ref{tab:ai_ready_comparison} summarizes the minimum transformation required to render such content machine-consumable.

\begin{table}[htbp]
\centering
\caption{Package of Practices: Comparison of Current and AI-Ready Formats}
\label{tab:ai_ready_comparison}
\small
\begin{tabular}{p{2.5cm}p{3cm}p{5.4cm}}
\toprule
\textbf{Aspect} & \textbf{Current} & \textbf{AI-Ready} \\
\midrule
Format & PDF / HTML documents & JSON or XML with schema \\
Identifiers & Crop/variety names in text & Standardized crop, variety, and agro-ecological codes \\
Timing & Relative phrases (e.g., ``early season'') & ISO 8601 dates, DAS\textsuperscript{*}-based stages \\
Quantities & Prose with mixed units & Numeric values with SI units and unit codes \\
Pesticide info & Product trade names in tables & Active ingredients with CAS numbers, dosages \\
Fertilizer dosage & Narrative recommendations & Structured nutrient amounts (N-P-K) with application schedules \\
Irrigation & Descriptive schedules & Timing, volume, and soil moisture thresholds \\
Logic \& conditions & Narrative conditions & Explicit IF-THEN rules with decision parameters \\
Access & Manual download & API-based queries with filters \\
Versioning & Publication year only & Versioned, time-stamped, auditable records \\
\bottomrule
\end{tabular}
\vspace{2mm}

{\scriptsize \textsuperscript{*}DAS = Days After Sowing; CAS = Chemical Abstracts Service}
\end{table}

\subsection{Illustrative Case: Integrated Crop Area Estimation}

\subsection{Minimum Schema Requirements}

An AI-ready PoP requires a limited set of core elements: persistent identifiers, explicit temporal and spatial context, quantitative precision, and machine-readable conditional logic. To enhance interoperability across agencies and states, data schemas should align with existing institutional taxonomies rather than introduce parallel standards \cite{arnaud2020ontologies}. 

For instance, crop and variety identifiers should align with ICAR’s standardized vocabularies, while agro-ecological contexts should be mapped to established spatial frameworks like the NBSS\&LUP zoning.

Each PoP record should also be persistently identifiable and versioned. Assigning a stable \texttt{pop\_id} through persistent resolvers (e.g., Handle or DOI systems) enables reproducibility by allowing models to reference immutable snapshots of recommendations as they evolve over time.

\subsection{Transformation Effort}

Recent open-source toolkits such as Docling \cite{livathinos2025docling} provide layout-aware PDF-to-JSON pipelines capable of recovering tables, headings, and reading order with high token-level fidelity on scientific documents. However, their performance on Indian PoP narratives, which often mix prose, tables, and region-specific conventions, remains unbenchmarked. Conversion accuracy, validation effort, and error rates therefore remain open empirical questions.

\subsection{Why This Matters Beyond PoP}

Although illustrated using PoPs, the same data-readiness constraints apply across agricultural domains, including soil health records, weather advisories, pest surveillance, and market intelligence. In the absence of structured schemas, persistent identifiers, and APIs, AI systems expend disproportionate effort on data cleaning rather than inference. Establishing AI-ready data foundations therefore reallocates effort from bespoke pre-processing toward scalable model development and deployment\cite{chatterjee2019crm}.

In this sense, AI readiness is not an algorithmic property but a data-infrastructure condition. Addressing it is a prerequisite for moving beyond pilot projects toward reliable, transferable, and equitable AI use in Indian agriculture.

\section{ Digital Infrastructure Reforms in Indian Agriculture}\label{sec6}

India’s agricultural digital ecosystem is evolving toward coordinated national and state platforms that support geospatial analytics, federated registries, and data exchange. This section reviews key initiatives based on their scope, operational status, and relevance for data-driven and AI-enabled applications.

\subsection{National Initiatives}
\subsubsection{Krishi-Decision Support System (Krishi-DSS)}

The Krishi-Decision Support System (Krishi-DSS) was launched nationally on 16 August 2024 as a geospatial decision-support platform integrating satellite imagery, crop survey outputs, soil indicators, weather data, and water resources information. It is intended to support near-real-time monitoring and advisory functions for farmers, extension systems, and policymakers \cite{samudra2025geospatial}.

Krishi-DSS consolidates datasets that were previously dispersed across agencies, strengthening national-level coordination in crop assessment, drought and water monitoring, and seasonal planning. It is positioned as a core analytical component of India’s Digital Public Infrastructure for Agriculture. At present, the platform primarily delivers dashboards and analytical outputs through government channels, while publicly documented programmatic access for external developers remains limited \cite{krishidss2024}.

\subsubsection{AgriStack (Digital Public Infrastructure for Agriculture)}

AgriStack provides the foundational framework for federated agricultural data registries, including farmer identities, geo-referenced land parcels, and crop-sown records. The architecture emphasizes interoperability, consent-based data sharing, and state-level data sovereignty.

Implementation is proceeding in phases across states, with progress closely tied to the digitization and validation of land and farmer records. While policy documents outline standard identifiers and service interfaces, operational maturity varies across states. AgriStack’s principal contribution to date is establishing a common architectural direction for future integration and reuse of agricultural data systems.

\subsubsection{Unified Portal for Agricultural Statistics (UPAg)}

The Unified Portal for Agricultural Statistics (UPAg) \cite{UPAgc1} serves as a centralized platform for compiling and publishing agricultural statistics and operational crop estimates from multiple agencies. It improves accessibility and comparability of official statistics through dashboards and summary indicators.

UPAg currently prioritizes consolidated reporting and visualization. Publicly available materials indicate limited support for large-scale machine-readable data access or automated integration into analytical pipelines, positioning the platform primarily for human-led analysis and policy monitoring.

The relative maturity, scope, and accessibility of these digital agriculture initiatives are summarized in Table~\ref{tab:digital_agri_infra}

\begin{table}[htbp]
\centering
\caption{Status of Major Digital Agriculture Infrastructure Initiatives in India}
\label{tab:digital_agri_infra}
\small
\begin{tabular}{p{1.5cm} p{1cm} p{3cm} p{2cm} p{2.8cm}}
\hline
\textbf{Initiative} & \textbf{Level} & \textbf{Function} & \textbf{Status} & \textbf{Data  Access} \\
\hline
Krishi-DSS & National & Geospatial analytics and decision support using satellite, weather, soil, and water data & Operational & Public API documentation not available; data accessible only through MoA dashboards \\
\hline
AgriStack & National & Federated registries for farmers, land parcels, and crop data & Phased rollout across states & Access policy differs across states; no unified open-API framework published \\
\hline
UPAg & National & Consolidation and publication of agricultural statistics and crop estimates & Operational & Data accessible via interactive dashboards; machine-readable APIs not offered \\
\hline
ADeX (Telangana) & State & Standardized agricultural data exchange with consent-based access & Operational & REST APIs documented; access subject to data-user agreement with Government of Telangana. \\
\hline
VISTAAR & National & Advisory dissemination and information aggregation & Operational & No public API documentation; data released as advisories \\
\hline
\end{tabular}
\vspace{1mm}

\footnotesize
\textit{Source: Compiled from various sources \cite{Beriya2022,adex2023,WEF2024_agriDPI,UPAgAPI2025,krishidss2024}.}
\end{table}

\subsection{State-Level Digital Data Platforms}
This subsection highlights selected state-level initiatives that illustrate distinct approaches to agricultural data infrastructure.

\subsubsection{Telangana Agricultural Data Exchange (ADeX)}

Telangana’s Agricultural Data Exchange (ADeX) \cite{adex2023} represents the most advanced state-level agricultural data platform in India. Built on data exchange principles, it provides standardized, machine-readable datasets through documented APIs and incorporates consent-based access controls supported by a formal Agriculture Data Management Framework.

ADeX hosts datasets spanning soil health, weather, crop yields, irrigation infrastructure, and market prices, and has supported multiple applied use cases in advisory services and market intelligence. Its operational status demonstrates the feasibility of API-enabled agricultural data exchange within a state context, though replication elsewhere depends on institutional capacity and sustained investment.

\subsubsection{Karnataka Farmer Registration \& Unified ID for Traceability System (FRUITS)}

Karnataka's FRUITS platform establishes a foundational digital registry \cite{FRUITS_Karnataka} of 10 million farmers, linking land records, crop data, and service entitlements through a unified ID system. Unlike data exchange platforms, FRUITS focuses on digital identity and service delivery integration across state departments, enabling targeted input subsidies, crop insurance, and extension services. Its integration with Bhoomi land records demonstrates practical inter-operability within state systems, though data standardization and update frequency remain ongoing challenges.

\subsubsection{Other State Initiatives}

Maharashtra's MahaAgriTech integrates satellite imagery with ground-truth data for crop monitoring and insurance, while Andhra Pradesh's Rythu Bharosa Kendras aggregate farmer data for scheme disbursement. These initiatives operate primarily as internal administrative systems rather than open data platforms, reflecting a narrower focus on governance efficiency over ecosystem-wide data sharing \cite{saifuddin2023effectiveness}.

\subsection{Advisory and Information Platforms}

India’s digital agriculture landscape also includes platforms focused on advisory dissemination and information aggregation, such as Virtually Integrated System to Access Agricultural Resources (VISTAAR). These systems play an important role in outreach and service delivery but are not designed as structured data infrastructures. Their outputs are primarily textual or semi-structured, limiting direct applicability for advanced analytics without additional transformation.

\subsection{Synthesis}

\begin{figure}[htbp]
\centering
\includegraphics[width=0.8\textwidth]{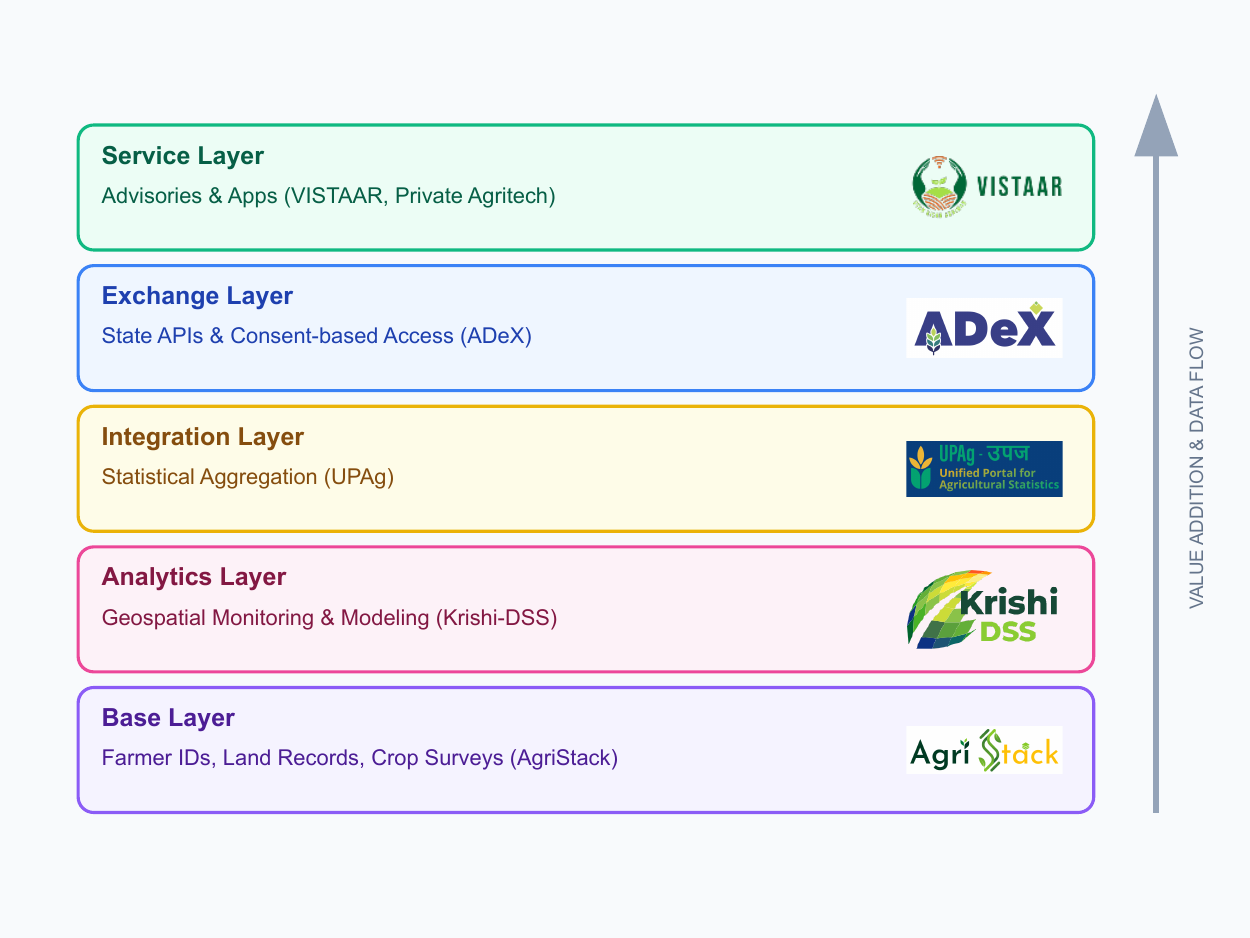}
\caption{Layered architecture of digital agriculture infrastructure in India}
\label{fig:digital_agri_arch}
\end{figure}

Fig.~\ref{fig:digital_agri_arch} illustrates the layered structure of India’s emerging digital agriculture infrastructure, showing how national analytics, federated registries, state exchanges, and service platforms interact.

India’s digital agricultural infrastructure now comprises multiple functional layers: national geospatial analytics (Krishi-DSS), federated registry frameworks (AgriStack), statistical consolidation portals (UPAg), and state-level data exchanges (ADeX). Together, these initiatives indicate a transition toward a more coherent digital ecosystem. Current strengths lie in government-led monitoring and state-bounded applications, while broader AI-driven scalability depends on continued progress in standardization, interface maturity, data quality, and institutional capacity.

\section{Private-Sector Perspectives and Collaborative Models}
\label{sec7}
\subsection{Data access}
Private developers must obtain public agricultural micro-data through channels whose availability, usage conditions, and coverage are not always disclosed in advance; this can lengthen early-stage planning and product-development cycles \cite{wef_future_farming_2025}.
\subsection{Alternative data strategies}
Many ventures bypass administrative datasets by combining open satellite imagery, crowd-sourced field observations and anonymised transaction records from farmer co-operatives. While these approaches maintain development momentum, this sacrifice granularity-limiting the precision of soil health advisories, yield forecasts and risk models that detailed cadastral, registry and survey data could enable.
\subsection{Structured access frameworks}
Some states are testing rule-based protocols:
Telangana's ADeX, launched in August 2023, provides consent-based API access to anonymised farmer registry and crop-sown data under published tier pricing and automated usage logging \cite{wef_future_farming_2025,adex2023}.
These frameworks aim to balance private sector needs for predictable access with public oversight through explicit terms, consent mechanisms and audit trails.
\subsection{Policy implications}
Where access rules are transparent, private capital can shift from data substitution toward collaborative reuse-potentially accelerating innovation cycles. Where uncertainty persists, investment may flow toward open-data alternatives, shaping both market structure and the technical architecture of emerging agricultural platforms.
Clarifying data licensing frameworks could strengthen digital public infrastructure by enabling more effective public-private collaboration.

\section{AI Applications Enabled by Robust Agricultural Data Systems}\label{sec8}

\begin{figure}[htbp]
\centering
\includegraphics[width=0.9\textwidth]{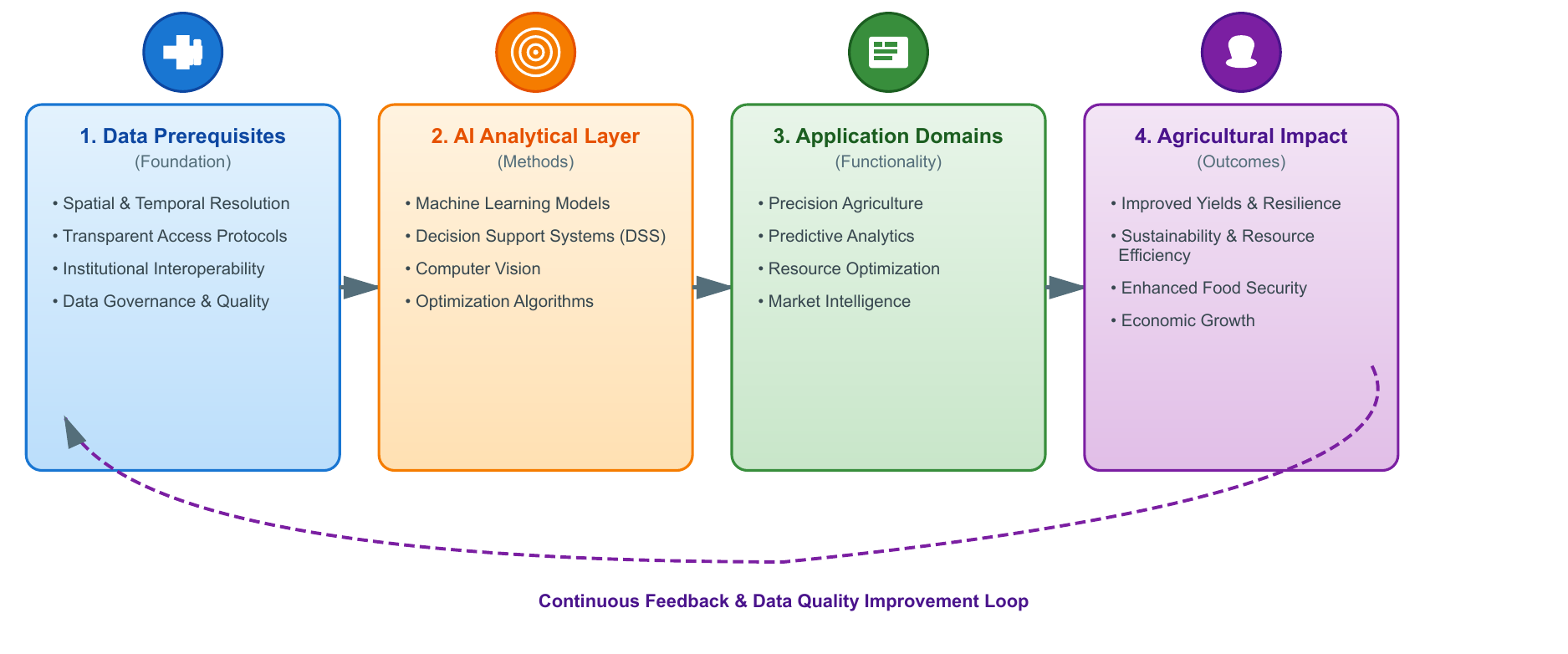}
\caption{AI application framework: from data systems to agricultural impact}
\label{fig:ai_framework}
\end{figure}

Robust and interoperable agricultural data systems constitute a foundational prerequisite for the effective deployment of AI in agriculture. When data are available at appropriate spatial and temporal resolutions, governed by transparent access protocols, and interoperable across institutional and technical boundaries, AI methods can be applied to a wide range of analytical and decision-support tasks. Accordingly, this section reviews major AI application domains enabled under such data conditions—precision agriculture, predictive analytics, resource optimization, and market intelligence \cite{shekhar2017agriculture}. As illustrated in Fig.~\ref{fig:ai_framework}, these applications emerge through a layered pipeline linking data foundations to analytical methods and downstream agricultural impacts, with performance fundamentally conditioned by upstream data quality, continuity, and governance.

\subsection{Core Application Domains}

Across domains, AI capabilities emerge as \emph{system-level functions} enabled by integrated data infrastructures rather than as standalone algorithmic solutions. Differences in performance across applications can largely be attributed to variation in data coverage, resolution, and interoperability.

\textbf{Precision Agriculture.}  
Precision agriculture applications use spatially explicit data to tailor crop management practices within and across fields. Typical use cases include variable-rate input application, crop stress detection, and yield mapping. These applications rely on the joint availability of remote sensing imagery, soil and terrain datasets, localized weather observations, and farm management records. Evidence from the literature indicates that effectiveness depends less on algorithmic sophistication than on dataset consistency \cite{Barsi02102019}, georeferencing accuracy, and temporal continuity.

\textbf{Predictive Risk Analytics.}  
Predictive analytics support anticipatory decision-making by estimating the likelihood of adverse outcomes such as yield shortfalls, pest outbreaks, or weather-induced stress. Such systems integrate long historical records with near-real-time observations, combining agronomic, meteorological, and environmental variables. Prediction reliability is strongly influenced by the length of historical records, treatment of missing data, and alignment between modeled variables and the spatial and temporal scales at which decisions are made.

\textbf{Resource Optimization.}  
AI-enabled optimization aims to improve the efficiency of water, nutrient, and energy use while maintaining productivity. Operating at fine temporal scales, these applications require timely data on weather conditions, soil moisture status, crop growth stages, and management history. Sensor gaps, reporting delays, or coarse spatial aggregation substantially degrade performance, underscoring the importance of reliable, low-latency data pipelines.

\textbf{Market Intelligence.}  
Market-oriented applications support functions such as price discovery, demand–supply forecasting, and marketing-timing decisions. These systems typically integrate price series, market arrivals, trade flows, and exogenous drivers, including weather shocks and policy interventions. In this domain, system value is closely linked to data latency, cross-market consistency, and transparency in data revision practices—factors that are especially critical in environments characterized by pronounced information asymmetries. In practice, achieving low data latency, cross-market consistency, and transparent revision protocols remains challenging, particularly in settings marked by fragmented data systems and institutional heterogeneity.

\subsection{Agricultural Statistics and Policy Intelligence}

While precision agriculture focuses on field-level optimization, national agricultural statistics support macro-level functions such as production forecasting, market planning, food security monitoring, and trade policy. Both rely on multi-source data integration, but agricultural statistics operate at coarser spatial scales (district to national) and seasonal temporal resolution.

\begin{figure}[htbp]
\centering
\includegraphics[width=0.8\textwidth]{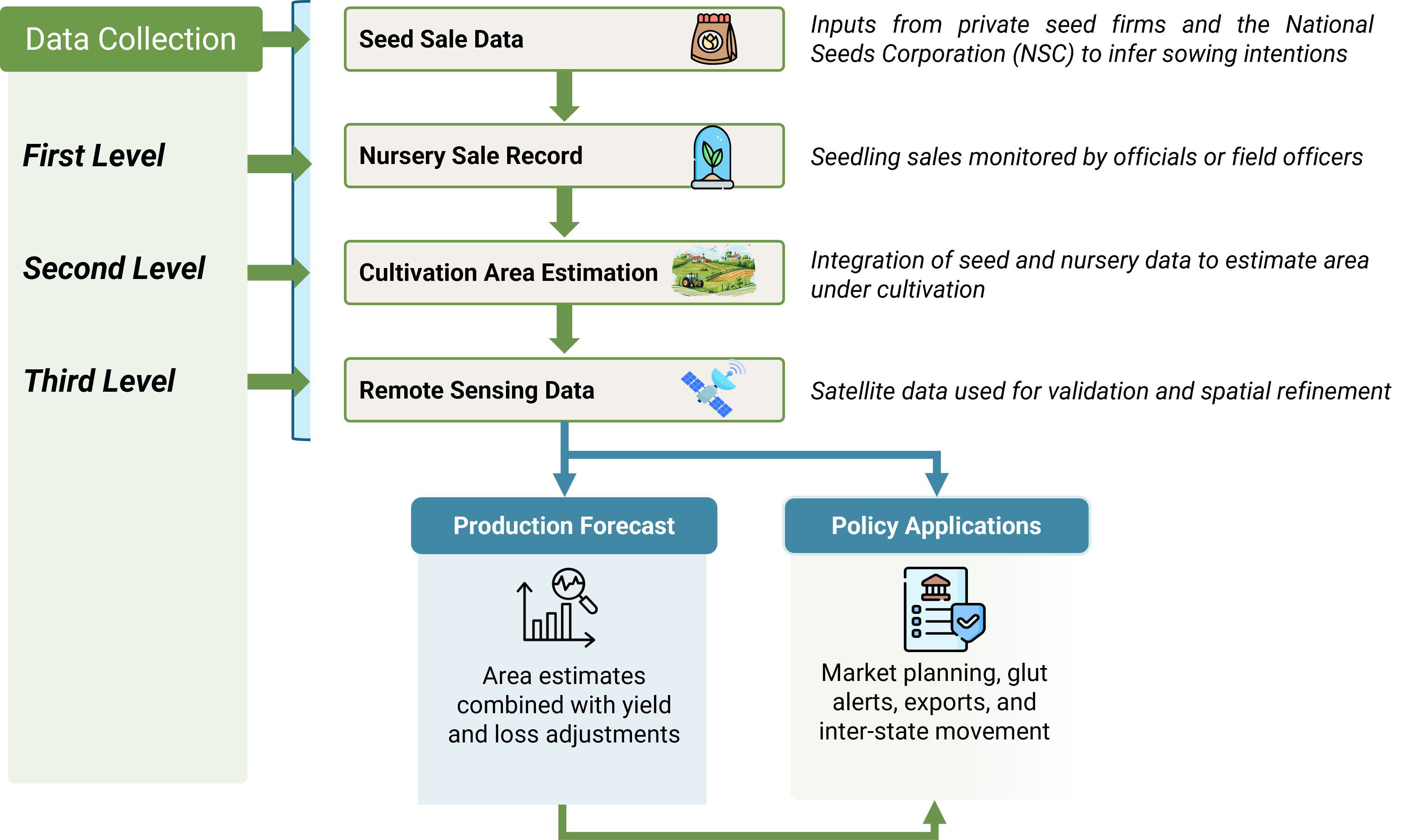}
\caption{Integrated Framework for Crop Area Estimation}
\label{fig:croparea}
\end{figure}

Fig.~\ref{fig:croparea} illustrates an integrated framework for AI-enabled crop area estimation. First-level inputs combine private-sector seed sales with public inventory records (NSC) to infer prospective sowing intentions. Second-level verification incorporates nursery sale records monitored by field officers and their consolidation into cultivation area estimates. Third-level validation uses remote sensing data to verify the realized area under cultivation against reported intentions. The validated area estimates, when combined with yield and loss adjustments, form the basis for production forecasts that support downstream policy applications such as market planning, export decisions, and interstate logistics.

The framework highlights core AI-readiness requirements discussed in Section~\ref{sec5}: the availability of persistent spatial identifiers, machine-readable data formats enabling interoperability between public and private data sources, temporal alignment between input sales and crop cycles, and consistent semantic definitions of cultivated area across administrative and remote sensing datasets. In this sense, the framework is conceptually aligned with the Agri-BIGDATA integration model proposed by Yang et~al.~ \cite{YANG2020150}, which emphasizes structured fusion of administrative, biological, and remote sensing data for decision support.

Overall, the framework clarifies how administrative input data, field-level verification, and remote sensing can be systematically integrated into a unified pipeline for crop area estimation and policy-relevant production intelligence.

\subsection{Data Requirements and Integration Challenges}

\begin{table}[htbp]
\centering
\caption{AI Applications: Essential Data Streams and Quality Requirements}
\label{tab:data_requirements}
\renewcommand{\arraystretch}{1.15}
\small
\begin{tabular}{p{2cm} p{2cm} p{3cm} p{3cm}}
\hline
\textbf{Application} & \textbf{Primary Data Needs} & \textbf{Spatial / Temporal Scale} & \textbf{Indicative Quality Requirements} \\
\hline
Precision Agriculture & Remote sensing, soil data, weather, field records & Sub-field to field; weekly & Stable georeferencing, low cloud contamination \\
Predictive Risk Analytics & Historical yields, weather time series, vegetation indices & Regional; daily to seasonal & Long time series, limited missing values \\
Resource Optimization & Weather forecasts, soil moisture, crop stage data & Field-level; daily & Low latency, sensor reliability \\
Market Intelligence & Prices, arrivals, trade indicators & Market-level; daily & Timely updates, consistent reporting \\
Climate Impact Modeling & Climate projections, crop parameters & Regional; daily & Bias-corrected, downscaled inputs \\
Agri-finance Applications & Remote sensing, farm activity records & Farm-level; seasonal & Ground-truth validation of indices \\
Policy making, planning and monitoring & Remote sensing, official statistics & District to national; multi-year & Temporal consistency and cross-source harmonization \\
\hline
\end{tabular}
\end{table}

Across application domains, common integration challenges persist \cite{devare2023governing,coble2018big}: heterogeneity in data formats (e.g., raster versus vector), misaligned spatial and temporal scales, limited interoperability across proprietary systems, and the absence of standardized metadata. These frictions propagate throughout AI pipelines—from data ingestion to model deployment—and disproportionately constrain applications designed for smallholder-dominated agricultural systems (see Table~\ref{tab:data_requirements}).

\emph{Section-level synthesis.}  
Taken together, the reviewed evidence indicates that AI performance in agriculture is shaped less by algorithm choice than by the strength and coherence of the surrounding data ecosystem. Investments in data standardization, interoperable pipelines, and transparent governance therefore function as enabling infrastructure, determining which applications are technically robust, economically viable, and scalable in practice.

Consistent with this assessment, the literature documents substantial performance losses when models trained on regionally biased or unrepresentative datasets are deployed in new agro-ecological contexts \cite{noyan2022plantvillage,priyatikanto2023improving}, as well as systematic failures under distributional shifts induced by climate variability, pest evolution, or market shocks \cite{rabanser2019failing}. Consequently, within the broader review narrative, this section reinforces the conclusion that the binding constraint on AI utility in Indian agriculture lies not in algorithm availability but in the inclusiveness, continuity, and governance of the data systems on which these applications depend.

\section{International Lessons and Transferable Insights}\label{sec9}

\begin{figure}[htbp]
\centering
\includegraphics[width=0.8\textwidth]{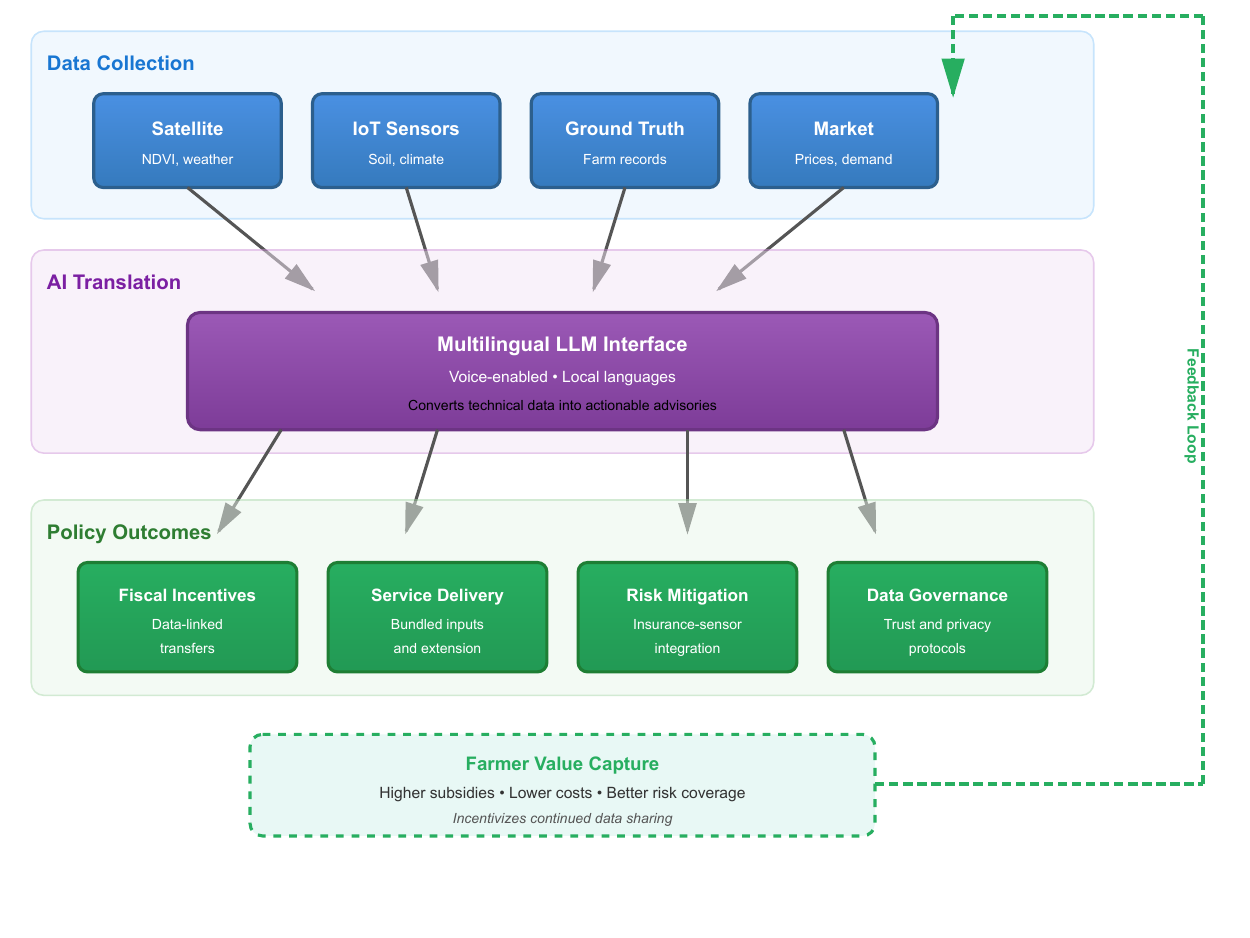}
\caption{Conceptual framework for multilingual AI-driven agricultural advisory infrastructure}
\label{fig:agri_value}
\end{figure}

To translate global experiences into the Indian context, we categorize international evidence into four thematic pillars. Each pillar represents a critical ``design rule'' for building an AI-ready agricultural stack that accounts for smallholder constraints and diverse agro-climatic conditions. As illustrated in Fig.~\ref{fig:agri_value}, these pillars jointly define a data-to-value chain in which interoperable data systems, AI translation layers, governance mechanisms, and farmer-facing interfaces interact to generate measurable agronomic and policy outcomes.

\subsection{Pillar 1: Data-Contingent Fiscal Transfers}
The EU's \textit{Integrated Administration and Control System (IACS)} illustrates that digital adoption is most effective when structurally linked to financial incentives. By tying subsidy eligibility to georeferenced parcel data and satellite-derived NDVI, the framework achieved a 96~\% reduction in inspection costs (\texteuro{}60--70 per remote check vs. \texteuro{}1,800 for physical visits) while improving targeting precision \cite{niassembly2015}.

\textit{For India, this translates directly into a policy imperative:} the country should evolve its income-support frameworks toward a ``Conditional DBT'' model. A designated portion of central transfers could be framed as a \textbf{Data-Utility Bonus}, released only when ground-truth sowing data or geotagged assets are validated through the national digital architecture.

\subsection{Pillar 2: Bundled Service Delivery}
Evidence from China’s decade-long agricultural extension reforms indicates that data analytics generate the greatest value when bundled with physical inputs and localized expertise. The integration of coordinated soil-testing databases with village-level extension networks supported a ten-year intervention reaching over 21 million farmers, resulting in yield increases exceeding 10 percent alongside 15--18 percent reductions in nitrogen fertilizer use, with estimated collective savings of \$12.2 billion \cite{yale2018}.

\textit{Adapting this model to India requires a structural shift:} precision-ag interventions must be channeled through \textbf{FPO-led service bundles}. Legal and financial incentives for equipment subsidies should be contingent on mandatory soil-testing and buy-back agreements to ensure data-driven advice leads to measurable productivity gains.

\subsection{Pillar 3: De-risking via Sensor-Insurance Loops}
The Israeli model illustrates how high-quality farm data collection can be made affordable. By subsidizing data-linked insurance premiums, the state reduces the upfront cost of on-farm sensors and reframes them as tools for risk management. Reported trials show water-use reductions of about 40 percent in alfalfa without yield loss \cite{cropx2023}, while government statistics indicate that 68 percent of vegetable area is now covered by sensor-based insurance, with insurer loss ratios declining to 0.61 \cite{israel2023}.

\textit{India can operationalize this through an immediate policy modification:} the \textit{PMFBY} (or future insurance iterations) should introduce a \textbf{Sensor-Linkage Rebate}. Farmers providing real-time API access to approved soil-moisture probes receive an actuarial discount, creating a high-fidelity data loop for regional crop monitoring.

\subsection{Pillar 4: Instructive Failures in Governance and Trust}
Sustainable digital stacks require proactive management of fiscal and social risks. The \textbf{Malawi Farm Input Subsidy Programme (2011--2018)} demonstrates that adding e-voucher layers without integrating soil-test data increased administrative costs without measurable yield gains \cite{fao2019}. Furthermore, Organisation for Economic Co-operation and Development (OECD) analysis of EU agricultural data governance warns that broad third-party access without clear ``data-sovereignty'' protections creates trust deficits that hinder farmer participation \cite{oecd2020data}.

\begin{table}[h!]
\centering
\small
\caption{Thematic Synthesis of International Design Rules for India}
\label{tab:intl}
\begin{tabular}{p{3.2cm} p{4.2cm} p{5.7cm}}
\toprule
\textbf{Thematic Pillar} & \textbf{Benchmark Case} & \textbf{Design Rule for India} \\
\midrule
Fiscal Incentive 
& EU IACS Framework 
& Satellite-linked ``data bonus'' embedded within DBT structures \\[4pt]
Service Bundling 
& China Extension Reform 
& FPO-led subsidies tied to soil diagnostics and buy-back MoUs \\[4pt]
Risk Integration 
& Israel Agritech Ecosystem 
& Insurance premium rebates conditional on real-time API-based data sharing \\[4pt]
Governance Safety 
& Malawi / Netherlands 
& Prioritisation of real-time stock tracking and formal data-trust protocols \\
\bottomrule
\end{tabular}
\end{table}

Table~\ref{tab:intl} operationalizes the four thematic pillars by linking each to its empirical anchor and translating the design feature into the Indian institutional context.

\textit{Critically, translating these pillars to the Indian last-mile requires overcoming a significant ``cognitive and linguistic barrier.''} While the EU and Israeli models provide the administrative and technical logic, operational success in India will depend on accessibility. We propose the integration of Multilingual Large Language Models (LLMs) as the primary interface for the national Agri-Stack. By utilizing voice-enabled, local-language AI, the system can convert complex, AI-ready data into actionable advisories, bypassing literacy hurdles and ensuring that the digital stack is a bottom-up tool for the smallholder rather than a top-down monitoring device.

\section{Implementation Strategies and Recommendations}\label{sec10}

\begin{figure}[htbp]
\centering
\includegraphics[width=0.8\textwidth]{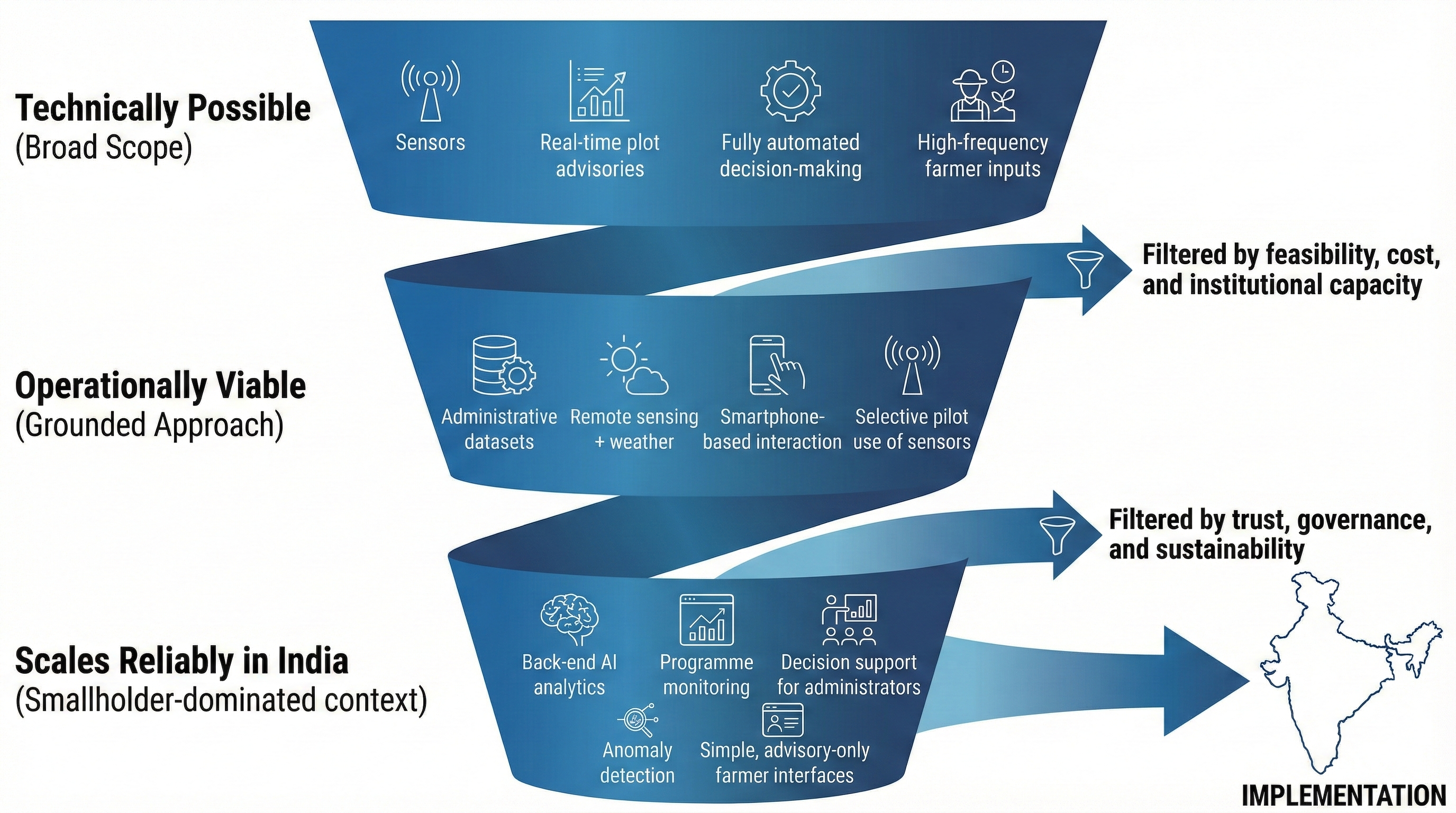}
\caption{Pathways to scalable AI implementation in Indian agriculture}
\label{fig:digital_agri_funnel}
\end{figure}

While Section~\ref{sec9} identifies transferable design principles from international experience, effective implementation in India depends on recognising operational boundaries and execution discipline. This section focuses on how digital and AI-enabled approaches can be deployed in ways that are institutionally feasible, fiscally realistic, and sustainable in a smallholder-dominated context. As summarized in Fig.~\ref{fig:digital_agri_funnel}, scalable AI deployment requires a sequenced approach that prioritizes data consolidation, institutional readiness, and selective use of analytics before expanding toward more advanced or farmer-facing applications.

\subsection{Pragmatic Use of AI}

AI deployment should prioritise use cases where analytical value is high and data demands are manageable. International experience shows that AI performs most reliably when applied to aggregation, pattern detection, and large-scale monitoring, rather than continuous, plot-level prescription across heterogeneous farming systems.

Accordingly, priority applications include programme monitoring, anomaly detection in crop and insurance datasets, and decision support for administrators and extension personnel. Farmer-facing AI advisories should remain limited, interpretable, and explicitly advisory, avoiding dependence on high-frequency inputs or hardware-intensive data streams.

\subsection{Sequencing and Institutional Readiness}

Implementation effectiveness depends critically on sequencing interventions in line with institutional capacity. Initial phases should focus on strengthening data consistency, validation mechanisms, and feedback loops within existing administrative workflows. Only after these foundations stabilise should more advanced analytical tools be introduced.

This incremental approach reduces the risk of underutilised pilots, mitigates implementation fatigue, and improves alignment with administrative realities and farmer trust.

\subsection{Data Quality and Accountability}

AI performance in agriculture is constrained less by model sophistication than by data quality. In the Indian context, inconsistencies in land records, crop declarations, and temporal alignment across datasets remain binding constraints.

Implementation should therefore prioritise reconciliation across sources, uncertainty flagging, and interpretability of outputs. AI-generated insights must be traceable and contestable to support administrative accountability and grievance redressal.

\subsection{Managing Hardware and Intermediation Dependence}

Sensor-intensive approaches face practical limits related to financing, maintenance, and uneven adoption among smallholders \cite{Ahmad2024AIAgriculture}. Sensors should therefore be treated as optional enhancers rather than core dependencies.

Similarly, while intermediary networks such as Common Service Centres (CSCs) can support onboarding and exception handling, routine reliance on them for advisory delivery risks introducing bottlenecks. Direct, smartphone-based interaction should be the default mode wherever feasible.

\subsection{Implementation Outlook}

The central implementation lesson is that digital agriculture in India will advance through consolidation rather than expansion. Systems that minimise hardware dependence, privilege direct farmer interaction, and use AI selectively for back-end analytics are more likely to scale and endure. Emphasising reliability, simplicity, and institutional fit offers a more credible pathway than replicating sensor-heavy or intermediary-driven models observed elsewhere.

\section{Conclusion}\label{sec11}
This review has examined the evolving role of data systems and artificial intelligence in agriculture, emphasizing that technological potential alone is insufficient to deliver meaningful transformation. Evidence from global experiences demonstrates that the effectiveness of AI-enabled agricultural solutions is fundamentally shaped by the quality of underlying data infrastructures, institutional arrangements, governance frameworks, and the extent to which technologies are adapted to local agro-climatic and socio-economic conditions.

By synthesizing international lessons and translating them into India-specific implementation strategies, the paper highlights that India’s pathway to digital agricultural transformation must prioritize adaptation over replication. Smallholder dominance, agro-climatic heterogeneity, and uneven digital access necessitate regionally calibrated AI models, publicly governed and interoperable data systems, and close integration of digital tools with extension and human institutions. Global failures-particularly those associated with data monopolization, pilot-driven fragmentation, and poor model transferability-offer valuable cautionary insights for avoiding similar pitfalls in the Indian context.

The analysis underscores that AI should be viewed not as a standalone solution but as a decision-support layer embedded within broader agricultural systems, complementing farmer knowledge, extension services, and policy interventions. Frugal innovation, explainability, trust, and inclusivity emerge as central design principles for ensuring widespread adoption and long-term sustainability.

Looking ahead, the success of AI in Indian agriculture will depend on coordinated investments in data infrastructure, institutional capacity, and governance mechanisms that align technological innovation with developmental objectives. If designed and implemented thoughtfully, AI-enabled systems can contribute to improved productivity, enhanced climate resilience, and more equitable outcomes for India’s diverse farming communities. However, realizing this potential requires sustained policy commitment, cross-sector collaboration, and continuous learning from both domestic experience and global evidence.

\section*{Acknowledgements}

The authors acknowledge the use of large language models (ChatGPT, OpenAI; Claude, Anthropic) for language editing, improving clarity of presentation, and assisting with high-level figure and diagram ideation. The responsibility for the content, analysis, interpretation, and conclusions remains entirely with the authors.

\section*{Conflict of Interest}
The authors declare that they have no competing interests.

\bibliographystyle{unsrt}  
\bibliography{references}

\end{document}